# Detection of Rotational Acceleration of Bennu using HST Lightcurve Observations

M. C. Nolan[1], E. S. Howell[1], D. J. Scheeres[2], J W. McMahon[2], O. Golubov[3,†], C. W. Hergenrother[1], J. P. Emery[4], K. S. Noll[5], S. R. Chesley[6], D. S. Lauretta[1]

[1]Lunar and Planetary Laboratory, University of Arizona, Tucson, AZ, USA. [2]Department of Aerospace Engineering Sciences, University of Colorado, Boulder, CO 80309, USA. [3]V. N. Karazin Kharkiv National University, 4 Svobody Sq., Kharkiv, 61022, Ukraine. [4]Earth and Planetary Science Department, Planetary Geosciences Institute, University of Tennessee, Knoxville, TN 37996, USA. [5]NASA Goddard Space Flight Center, Solar System Exploration Division, Greenbelt, MD 20770, USA. [6]Jet Propulsion Laboratory, California Institute of Technology, 4800 Oak Grove Drive, Pasadena, CA 91109, USA. [†]Also, Institute of Astronomy of V. N. Karazin Kharkiv National University, 35 Sumska Str., Kharkiv, 61022, Ukraine.

Corresponding author: Mike Nolan (nolan@orex.lpl.arizona.edu)

**Key Points:**

- Asteroid (101955) Bennu has been rotating faster over time since 1999.
- An acceleration in its rotation rate is needed to fit all the photometric measurements of asteroid Bennu.
- The measured acceleration of Bennu is consistent with theoretical calculations of the total effects of the YORP effect.

**Abstract**

We observed the near-Earth asteroid (101955) Bennu from the ground in 1999 and 2005, and with the Hubble Space Telescope in 2012, to constrain its rotation rate. The data reveal an acceleration of $2.64 \pm 1.05 \times 10^{-6}$ deg day$^{-2}$, which could be due to a change in the moment of inertia of Bennu or to spin up from the YORP effect or other source of angular momentum. The best solution is within 1 sigma of the period determined by Nolan et al. (2013). The OSIRIS-REx mission will determine the rotation state independently in 2019. Those measurements should show whether the change in rotation rate is a steady increase (due, for example, to the YORP effect) or some other phenomenon. The precise shape and surface properties measured by the OSIRIS-REx science team will allow for a better understanding of variations in rotation rate of small asteroids.

**Plain Language Summary**

We observed near-Earth asteroid (101955) Bennu—the target of the OSIRIS-REx spacecraft mission—using ground-based telescopes and the Hubble Space Telescope (HST). Our measurements show that its rotation has been speeding up since 1999. This change could be due to a change in the shape of Bennu or to the YORP effect: As the sunlight received by an asteroid is reflected or radiated back to space, the change in direction of the light coming in and going out pushes on the asteroid and can cause it to spin faster or slower, depending on its shape and orientation. The specific properties of the surface, such as the position of boulders, can have a large influence on the outcome. In the next few years, the OSIRIS-REx team will study Bennu and independently measure its rotation, helping us to better understand the structure and dynamical evolution of small asteroids.

# 1 Introduction

## 1.1 Approach to Bennu

Through late 2018, the Origins, Spectral Interpretation, Resource Identification, and Security–Regolith Explorer (OSIRIS-REx) spacecraft approached the near-Earth asteroid (101955) Bennu (hereafter, Bennu) to observe the surface characteristics and measure the spectral and thermal properties in detail (Lauretta et al., 2017). One of the reasons for choosing this target for the OSIRIS-REx mission is the wealth of information we have about its shape, albedo, and rotation rate based on extensive ground-based observations (Nolan et al., 2013; Hergenrother et al., 2013; Lauretta et al., 2015). Finding out how well our predictions match reality will add confidence to future interpretations of other objects. For these reasons, we determined the rotation rate and acceleration before arrival in December 2018.

## 1.2 Previous lightcurve observations and period comb

Nolan et al. (2013) determined the rotation rate of Bennu from shape modeling of radar and lightcurve observations in 1999 at the discovery apparition, and again in 2005 when Bennu made a close approach to Earth. The solution was ambiguous, allowing several integral rotations between these apparitions. The uncertainty is ± 5 rotations, giving a "comb," or a number of regularly spaced possible discrete solutions, for the rotation rate. This solution assumed a constant rotation rate.

Once the OSIRIS-REx spacecraft is in close proximity to Bennu, we will acquire imaging data to determine the rotation state. However, it is valuable to make our best effort using pre-approach data to predict the rotation rate and acceleration and then test those predictions. Spacecraft will never visit the vast majority of asteroids, and better interpretation of ground-based data is the only way to understand the population. By testing these predictions in a few cases, we will significantly improve our confidence in predictions made for other objects. Putting Bennu into context and constraining its history is also critical to optimize the analysis of the sample material brought back to Earth in 2023.

1.3 Resolving period ambiguities

Krugly et al. (2002) first measured the rotation period of Bennu using data from the 1999 apparition. They determined the period to be 2.1 h based on 3 days of optical lightcurve data. In that same apparition, Arecibo and Goldstone radar data suggested that the period was likely to be twice that long, discussed further by Hergenrother et al. (2013). Additional data in 2005 corroborated and refined the synodic period to $4.2905 \pm 0.0065$ h (Hergenrother et al., 2013). Shape modeling using a combination of radar and lightcurve data sets refined the sidereal period to a comb of solutions of $4.297461 \pm 0.00002$ h $\pm N \times 0.000352$ h, amounting to $12269 \mp N$ rotations over the 6-year interval between the observation sessions, where $N$ is an unknown small integer that we will determine in this paper.

Figure 1 shows uncertainty in the rotation rate schematically, with a comb of solutions separated by one revolution per 6 years. Each solution is individually uncertain by the ability to compare the phase of rotation of the model with the data (a few degrees) divided by the sampling interval (6 years). Our ability to compare the phase of rotation of the model with the data (still a few degrees) divided by the most extended period over which we know that we have counted all of the rotations (4 days) controls the overall envelope.

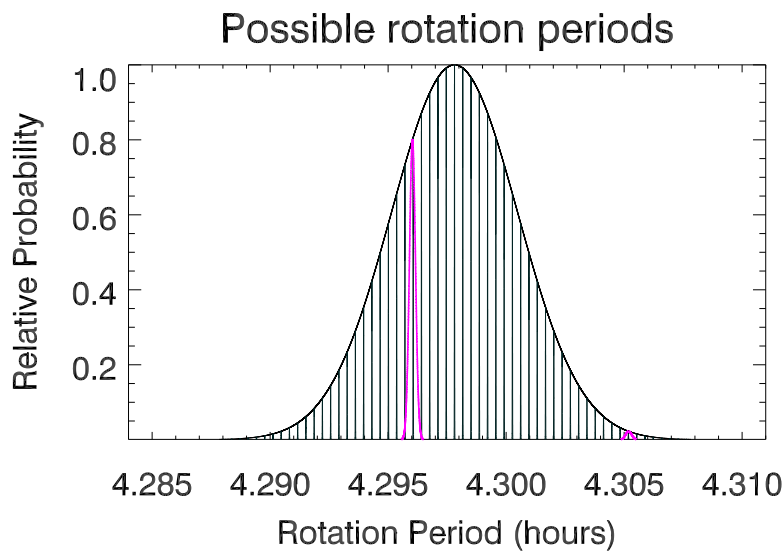

Figure 1. The rotation period of Bennu, showing the envelope from the 2005 lightcurve observations along with the comb of solutions (black vertical lines) allowed by combining the 1999 and 2005 data, as described in Nolan et al. (2013). The narrow pink curves near 4.296 and 4.305 hours were determined independently from the HST observations and show the +5 and

–21 periods discussed below. Each of the curves is plotted as a Gaussian with a sigma of 5 deg over the relevant sampling interval.

To distinguish between the allowable periods, we need additional phase information over a long enough interval that the resulting uncertainty in rotation period (the envelope in Figure 1) is less than the spacing between the tines of the comb.

The comb spacing is one full revolution over the 6 years between 1999 and 2005, or 0.164 deg per day. To distinguish between the tines of the comb, we require observations over a time interval T such that ~5 deg/T < 0.164, or T > 30 days.

## 2 Photometric Observations

### 2.1 Hubble Space Telescope observations

We observed Bennu with the Hubble Space Telescope (HST) on September 17 and 18 and December 10 and 11 of 2012. These observations reduced the rotation rate uncertainty so that we could determine an unambiguous period. In each of those epochs, we observed three consecutive orbits on one day followed by two orbits the following day to sufficiently sample Bennu's lightcurve given HST's orbital period around Earth.

We used the Wide-Field Camera (WFC3) with the F350LP filter, with 305s exposures in September and 306s in December, spaced ~6 minutes apart. To achieve that relatively high time resolution with HST, we used the WFC3 UVIS detector, reading out a 512-pixel subarray, and used no dithering, cosmic ray removal, or geometric correction. The standard HST cosmic ray removal process takes a set of images and median-filters them. We needed higher time resolution, so we took single images.

HST tracked the images at the asteroid's rate of motion. The images were processed using the standard pipeline by the Space Telescope Science Institute. We used IRAF to make the photometric measurements with a range of sky aperture sizes. The field was not particularly crowded, and only seven images were unusable because of Bennu being too close to a background star or galaxy, or irrecoverably contaminated by cosmic rays falling inside the aperture. We removed cosmic ray contamination from four images, and increased the uncertainties accordingly (column 5 of Data Set S1).

### 2.2 Absolute calibration

Using the HST zero point for the F350LP filter and appropriate aperture corrections, we obtained the absolute magnitudes for Bennu. We used an aperture with a 4-pixel radius to minimize the chances of cosmic ray contamination while still getting all of the Bennu flux. Several frames were measured out to a 10-pixel radius to determine the correction to the HST calibration standard. We also applied a color correction for the solar color at the effective wavelength of filter F350LP (5883.6 Å) interpolated from Johnson-Cousins BVR to be V – F350LP = –0.172 magnitudes (Dressel, 2018; Ramírez et al., 2012). The color for Bennu, determined from the spectrum (Clark et al., 2011) is +0.011 magnitudes. Data Set S1 gives the absolute magnitudes and the internal relative uncertainties. Our mean magnitudes in 2012 are about 0.09 magnitudes fainter than the apparent magnitudes calculated using the derived H and G values from Hergenrother et al. (2013). This difference is within 0.5 sigma of their estimated uncertainty in H of 0.20 magnitudes.

We use the shape model from Nolan et al (2013) (hereafter, the radar shape model) to generate synthetic lightcurves for comparison with the observations, using the illumination and viewing geometry from Bennu's ephemeris. The lightcurves in Figure 2 show December on the top and September on the bottom, shifted arbitrarily in phase to match the data. The difference in required phase shift from September to December (84 days) is 64.5 deg (plus or minus an integral number of rotations).

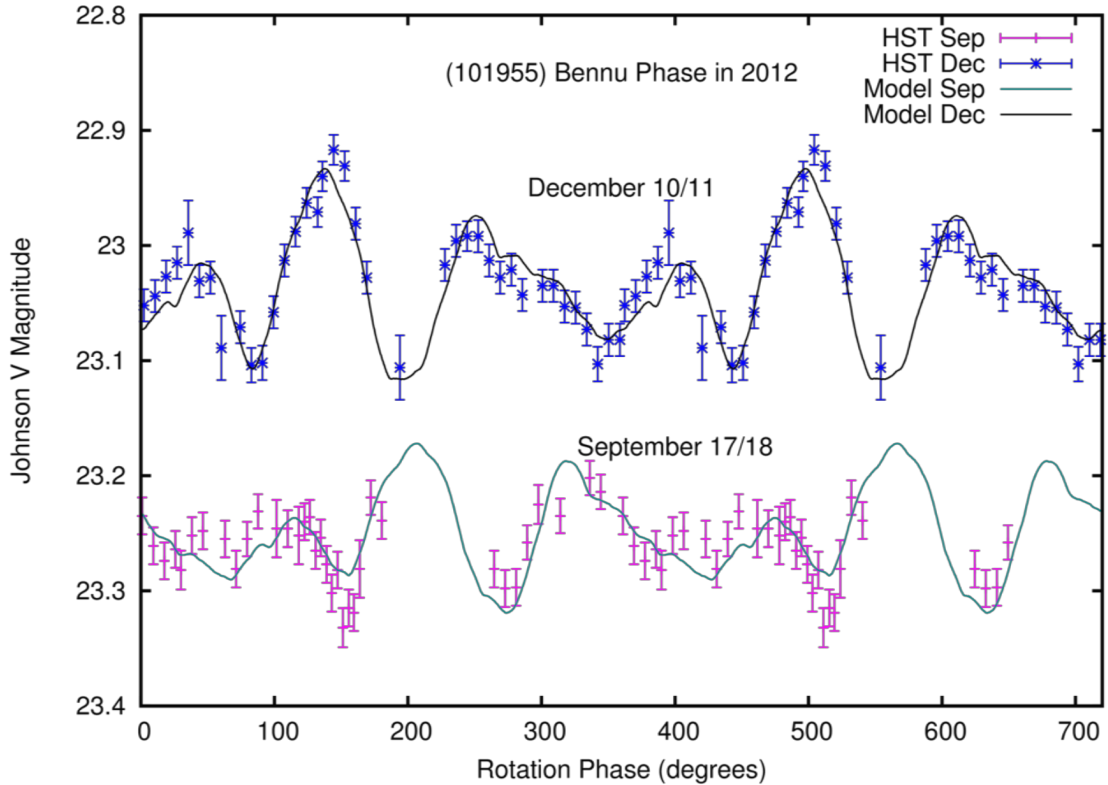

Figure 2. HST photometry phased to the rotation state from Nolan et al. (2013). December is on top, September below. Solid lines are the radar shape model predicted lightcurve for that date shifted separately to fit each of the observing sessions. The apparent horizontal shift between the two dates is 64.5 deg, indicating that the rotation period that Nolan et al. (2013) adopted is not precisely correct.

## 3 Lightcurve Fitting

### 3.1 Testing the comb of solutions to isolate the correct period

The 64.5 deg phase offset required to align the lightcurves between September and December gives a rotation rate of 64.5 deg/84 days = 0.768 deg day$^{-1}$ faster (just under 1 sigma) than the nominal period from Nolan et al. (2013), very nearly corresponding to the tine of the comb with 5 extra rotations. For completeness, we examined the entire comb of possible solutions ± 30 rotations (>5 sigma) from the nominal solution. Only the solutions at +5 and −21 rotations fit the HST observations by themselves (Figure 1, pink curves).

As with the two ground-based observing epochs, the two sessions of the HST observations constrain the period to lie on a comb of solutions, but with a different comb spacing than the ground-based lightcurves. The solution at −21 rotations has one fewer rotation between the September and December 2012 HST observations than the solution at +5.

We now consider the absolute rotation phase. Using the +5 solution, fit only to the 1999 and 2005 data, we show in Figure 3 that the 2012 data are nearly in phase (within about 15 deg), 14,289 rotations later. In contrast, the −21 rotation solution is ~180 deg out of phase. As it is also only consistent with the Nolan et al. (2013) results at the 4-sigma level, we do not consider it further.

The new best-fit single rotation rate is P = 4.2960505 ± 0.0000008 h.

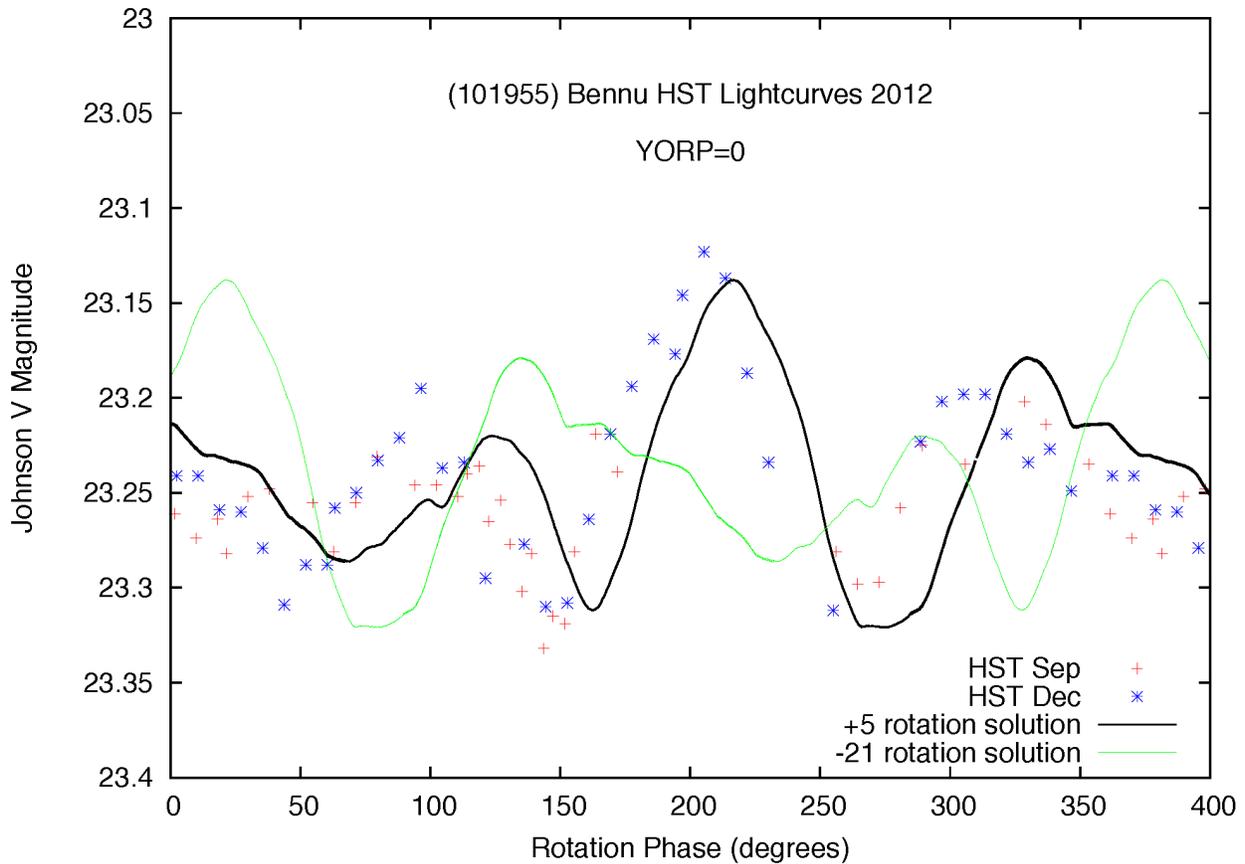

Figure 3. The 2012 HST data and model lightcurve are shown in terms of rotation phase. The +5 rotation solution shows rotation phases computed using a fixed rotation period of 4.2960505 h to fit the data in 1999 and 2005 (solid black curve). The model prediction for the −21 rotation solution is fit with a period of 4.3052265 h and is shown in green. The December data and model are shifted vertically by +0.21 magnitudes to match the September mean, error bars omitted for clarity. The phase offset between the +5 rotation model and the 2012 data is ~15 deg, but the −21 rotation solution is nearly 180 deg out of phase with the data.

3.2 Reducing residual phase error

Having found an unambiguous period, we attempted to refine the period using all three epochs (1999, 2005, and 2012) by allowing the pole orientation to shift slightly (within 15 deg). All of these attempts left a residual phase shift in the data. As an example, Figure 4 shows the residual phase error in 2005 when we fit the epochs 1999 and 2012. No single rotation rate can fit all of the data. The rotation rate appears to have changed.

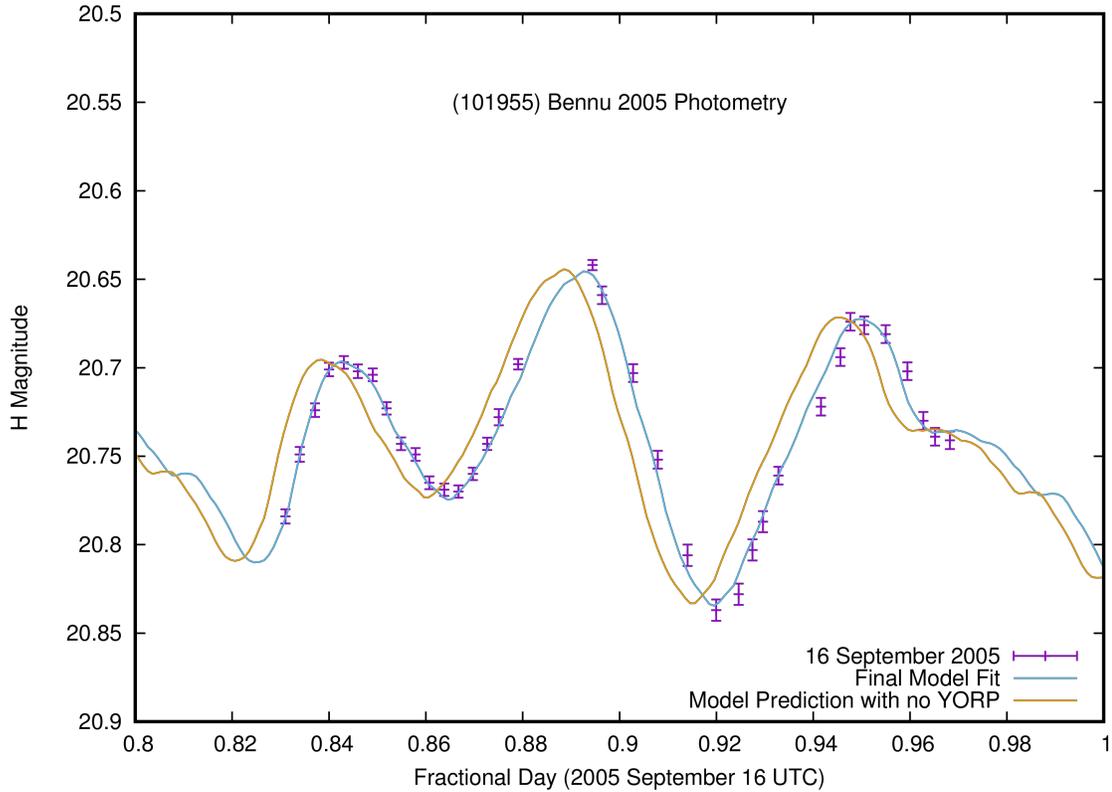

Figure 4. The solid gold curve uses the best acceleration-free model that fits the data in 1999 and 2012 to predict the lightcurve on 2005 September 16. The points are the data from Hergenrother et al. (2013) binned to 6 deg of rotation and are tabulated in Data Set S2. There is a visible phase shift of 0.005 days (9 deg) between the model and the data points. No single period can fit the lightcurve at all three observing epochs. The blue curve is the final model fit including a rotational acceleration of $2.64 \times 10^{-6}$ deg day$^{-2}$.

We added a constant angular acceleration to the model and found that the best-fit acceleration is $2.64 \pm 1.05 \times 10^{-6}$ deg day$^{-2}$. Bennu is rotating faster with time.

We compute $\chi^2$ as a function of the phase offset between the model lightcurves and the data. In each epoch, we assigned the phase uncertainty to be the difference for which reduced $\chi^2$ increased by 1 (Bevington, 1969). The phase uncertainties in 1999, 2005, and 2012 are 2.6, 1.4, and 5.0 deg, respectively. Some of the uncertainty is probably due to an imperfect shape model. We then computed the rotation state and uncertainty using a polynomial fit to these phases. The best-fit sidereal rotation rate of Bennu is $4.2960477 \pm 0.0000019$ h or $2011.1509 \pm 0.0009$ deg day$^{-1}$ on 2005 Sep 14.0 UTC with an acceleration of $+2.64 \pm 1.05 \times 10^{-6}$ deg day$^{-2}$.

## 4 Mechanisms of acceleration

There are several possible explanations for the change in rotation rate. Because we only have measurements at three epochs, a step change in spin rate cannot be ruled out. Changes to the spin rate from impacts are exceedingly unlikely during this 13-year interval. The maximum change in spin rate due to the 2005 planetary encounter with Earth is less than 5% of the uncertainty in our measurement (Scheeres et al., 2000). A change in the moment of inertia due to spin- or impact-induced reshaping could change the period, though the observed increase in rotation rate would require a contraction of the body, which would be an even more interesting scenario than a constant acceleration.

A plausible mechanism for the acceleration that has been observed on other near-Earth asteroids, beginning with (54509) YORP (Lowry et al., 2007, Taylor et al., 2007) is the Yarkovsky–O'Keefe–Radzievskii–Paddack (YORP) effect (Rubincam, 2000). This effect is a net torque that is present on an asymmetric asteroid that can either increase or decrease its rotation rate, depending on details of the rotational and, to some extent, the thermal properties (Vokrouhlický et al., 2015, and references therein).

Assuming that the increase in rotation rate is due to the YORP effect, we can compare our measurements to predictions of YORP acceleration. Using the radar shape model and associated pole and rotation rate, along with the homogeneous density estimate from Chesley et al. (2014), Scheeres et al. (2016) predicted the so-called Normal YORP acceleration for Bennu to be $-15 \times 10^{-6}$ deg day$^{-2}$. The Normal YORP effect only accounts for the photon emissions radiated normal to the surface (Vokrouhlický et al., 2015). This prediction is an order of magnitude larger and in the opposite direction of the observation reported here.

To better characterize the uncertainty in this value, we modelled the predicted Normal YORP effect for a suite of shapes similar to the radar shape model to compute the likely range of values using the computational model of McMahon & Scheeres (2010), which incorporates self-shadowing and secondary intersections of re-radiated energy. The shapes were changed by perturbing the vertices vertically using the random Gaussian spheroid method (Muinonen, 2010). We set the vertical perturbations to approximate the radar range resolution of 15 m, while we varied the correlation distances from 25 m (making small-scale, "spiky" topographic features) to 300 m (smoother global variations in topography). Figure 5 shows the range of the predicted Normal YORP effect from that suite of models. Our measured value of the acceleration lies within the distribution, but well away from the YORP prediction for the radar shape model shown by the heavy black dash-dot line. However, the distribution of computed YORP values is skewed toward positive values with respect to the previously computed radar shape model value.

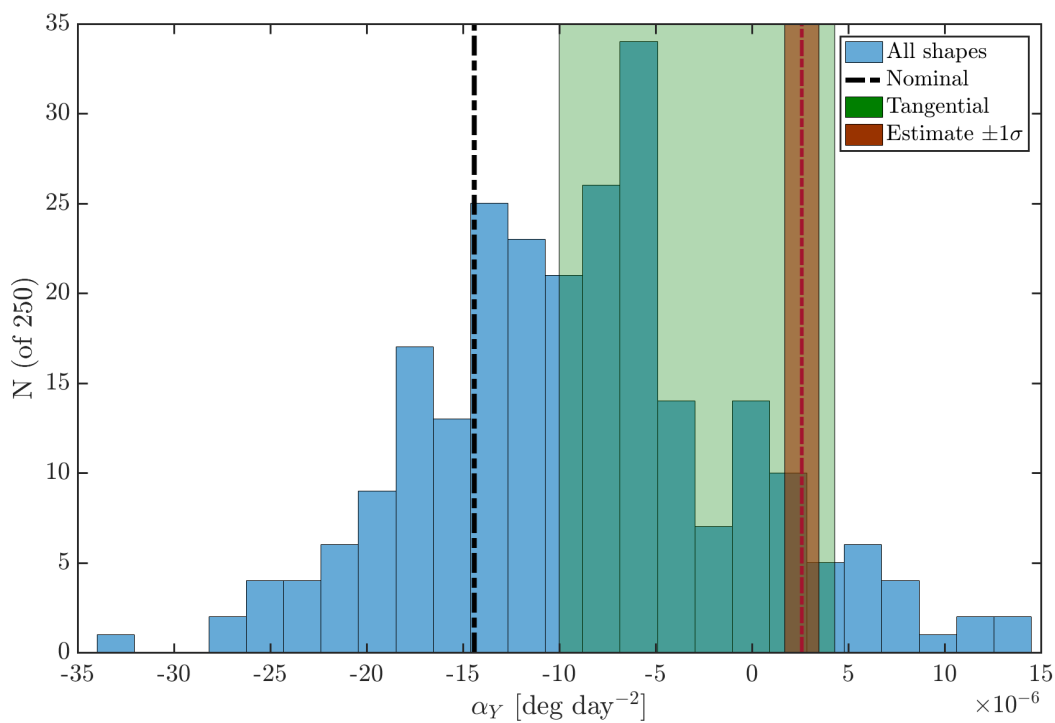

Figure 5. Shape model based YORP predictions for a population of 250 perturbed Bennu shapes. This set of shapes produces a mean YORP value of $-8.8 \pm 8.1 \times 10^{-6}$ deg day$^{-2}$. The estimated YORP value of $+2.64 \pm 1.05 \times 10^{-6}$ deg day$^{-2}$, shown by the red shaded region, falls within this distribution, though at a value greater than 92% of the tested shapes. The shaded green area shows the 1-sigma distribution of the combined predicted Normal and Tangential YORP accelerations using the radar shape and an Itokawa-like boulder distribution.

We also carried out a computation of the so-called Tangential YORP effect, following the methodology of Golubov et al. (2012, 2014). Tangential YORP arises from the solar heating of surface boulders and has a bias towards a positive torque, always accelerating a body's spin rate (as the sunset side of a boulder is warmer on average than the sunrise side). To carry out this computation, the size distribution of boulders on the asteroid must be specified, along with their heat capacity and thermal inertia. Across a range of expected values, and assuming a boulder distribution similar to Itokawa, the Tangential YORP contribution is estimated to be $+11.5 \pm 7.3 \times 10^{-6}$ deg day$^{-2}$.

When combined with the Normal YORP contribution from the radar shape model, the predicted total YORP acceleration ranges from $-10.3$ to $+4.3 \times 10^{-6}$ deg day$^{-2}$ (1 sigma). This is consistent with the currently measured acceleration and raises the prospects of validating the Tangential YORP theory.

**5 Discussion**

We anticipate that after detailed modeling of the shape, surface boulder covering, gravity, and thermal properties of Bennu, we will better understand the relative importance of these factors in determining the total YORP effect. Of particular interest is understanding the possible combination of Normal and Tangential YORP effects. The present shape of Bennu is similar to other objects near the rotational stability limit (Nolan et al., 2013; Scheeres et al., 2016), suggesting that perhaps it was rotating faster in the past. However, we find that it is accelerating.

At our measured acceleration rate, the mission science team should be able to measure the acceleration independently. At a resolution of 35 cm, detecting an offset of features by 0.08 deg of rotation should be possible after about 180 days of spacecraft observation. After more than 2 years, the detection will be extremely firm (McMahon et al., 2018). Knowing whether it is consistent over several decades will also be interesting. Any shifting of the surface features could affect the YORP rates. Scheeres et al. (2007) and Statler (2009) predict that even very slight shifts of surface material could change the acceleration rate or even reverse its direction. This prediction will allow us to test this idea over a timespan of a few decades.

## 6 Conclusions

Using the detailed rotation measurements in 1999, 2005, and 2012, we determined that a constant rotation rate does not fit the observations. An acceleration of $+2.64 \pm 1.05 \times 10^{-6}$ deg day$^{-2}$ is needed, which indicates that Bennu's rotation rate has increased over the past two decades. This measurement is opposite in direction and an order of magnitude smaller than the Normal YORP acceleration calculated by Scheeres et al. (2016) but is consistent with the Normal plus Tangential YORP contribution. It also lies within the distribution of Normal YORP predictions for 1-sigma perturbations of the radar shape model. The OSIRIS-REx science team will independently measure the rotational acceleration during its 2-years of proximity operations. The precise shape determination, surface boulder distribution, gravity measurements, and thermal property determinations will allow for a better connection between the dominant drivers of the YORP effect (if confirmed) and their relative importance. The OSIRIS-REx team can measure the stability of the rotation state, to confirm whether this acceleration is a steady increase due to the YORP effect, or some other (likely episodic) process such as mass movement. Thus, our observations form a critical baseline for future work.

## 7. Acknowledgements

We thank S. Lowry and an anonymous reviewer for helpful comments that improved this paper. Based on observations made with the NASA/ESA Hubble Space Telescope, obtained at the Space Telescope Science Institute, which is operated by the Association of Universities for Research in Astronomy, Inc., under NASA contract NAS 5-26555. These observations are associated with program 13118. The HST data are presented in Data Set S1, and the raw images are available from the Hubble Legacy Archive. The work of SRC was carried out at the Jet Propulsion Laboratory, California Institute of Technology, under a contract with the National Aeronautics and Space Administration. The work of JWM was supported by the National Aeronautics and Space Administration under Grant No. NNX14AN13G issued through the Planetary Geology and Geophysics Program. This material is based upon work supported by NASA under Contract NNM10AA11C issued through the New Frontiers Program.